\documentclass[a4paper,11pt]{article}
\pdfoutput=1
\usepackage{amsmath,amsfonts,hyperref,amssymb,empheq}
\usepackage{latexsym,mathtools,array}
\usepackage{physics}
\usepackage{bm}
\usepackage[dvipdfmx]{graphicx}
\usepackage[export]{adjustbox}

\makeatletter 
\newcommand{\LCASES}[1]{$\m@th\displaystyle{#1}$\hfil}
\newcommand{\CCASES}[1]{\hfil$\m@th\displaystyle{#1}$\hfil}
\newcommand{\RCASES}[1]{\hfil$\m@th\displaystyle{#1}$}
\makeatother
\newcases{ecases}{\quad}{\CCASES{##}}{\LCASES{##}}{\lbrace}{.}
\newcases{ecases*}{\quad}{\CCASES{##}}{{##}\hfil}{\lbrace}{.}

\begin{document}
\title{Hierarchical Autocatalytic Systems as a Bridge between Maximum Entropy Production and Bayesian Posterior Contraction: A Numerical Study with Stochastic-Thermodynamic Bounds}
\author{Yoshinori Watanabe \\ \href{mailto::xiangzey@gmail.com}{xiangze@gmail.com}}
\date{\today}
\maketitle

\begin{abstract}
We construct a three-layer reaction-diffusion model of an autocatalytic
chemical system in which raw molecules ($a_i$), catalytic proteins
($p_l$) and large RNA/protein ``genes'' ($W_p^{(k)}$) interact
through a mass-action stoichiometry tensor $\mathrm{Coef}_{ijk}$ whose
magnitude is modulated in $[-1,1]$ by the fold-stable activity
$\mathrm{pa}_l = \tanh(\beta W_{l,:,k}\, p_p - \theta_l)$ of the
largest polymers. Mass-action is broken by an $\varepsilon$-noise term
so that the system is genuinely nonequilibrium.

The model is implemented in NumPy/PyTorch and analysed through an
Optuna-based Bayesian search for sustainable parameter regions, after
which we compute (i) the total entropy production $\sigma(t)$, (ii)
the genetic Shannon entropy $S_\mathrm{gene}$ derived from a
symbol-string projection of the fold tensor, and (iii) the thermodynamic
uncertainty relation (TUR) and thermodynamic speed limit (TSL) bounds on
growth and evolution rates.

The hierarchical model exhibits the expected co-occurrence of
$\sigma_\mathrm{env}\!\uparrow$ and $S_\mathrm{gene}\!\downarrow$
predicted by Schrödinger's negentropy argument and reformulated as
maximum-entropy-production-principle (MEPP)-driven adaptation, with a
quantitative cost of $\sim 589$ nats of environmental entropy per nat
of genetic order.

However, the realised TUR product sits $10^4$--$10^5$ above the
universal bound of 2, and the TSL ratio sits $10^6$--$10^8$ above
its bound of 1. We trace the looseness to the multi-cycle structure of
the network and, by collapsing the system to a single
kinetic-proofreading-like cycle (a minimal Gillespie replicator), we
recover TUR products of $\sim 5$, matching the experimentally reported
regime of the ribosome \cite{PhysRevE.101.022415} . Scaling $(N,M,L)$ from
$(4,3,3)$ to $(32,8,3)$ leaves the looseness intact for the
hierarchical model but tightens it monotonically with particle number
for the minimal model.

We close by drawing an explicit correspondence between the autocatalytic
system and diffusion-model training: $a_\mathrm{ext} \to a$ flux $ \Leftrightarrow $ data-information flow, $ \tanh(\beta Wp - \theta) $ $\Leftrightarrow$ score network, replication noise $\Leftrightarrow$ forward-diffusion noise, $ S_\mathrm{gene} \searrow $ 
$\Leftrightarrow$ $ H[q(\theta|\mathcal{D})] \searrow $, and we discuss the implication
that biological cells operate far from thermodynamic optimality for 
reasons identical to why over-parameterised neural networks operate in
the ``lazy'' training regime. All code and figures are available as
supplementary material.
\end{abstract} 

\section{Introduction}
The thermodynamic basis of life has long been intertwined with the
notion of \emph{negentropy} introduced by \cite{WhatisLife}: living systems maintain internal order by exporting entropy to their
environment. Two modern frameworks have made this idea quantitative. The maximum entropy production principle (MEPP) \cite{10.1098/rstb.2009.0295}\cite{e27040449} posits that nonequilibrium systems
preferentially occupy macrostates that maximise entropy production rate,
and Dissipative Adaptation theory \cite{England2015Dissipative} shows that driven
self-organising systems preferentially populate
work-absorption-effective configurations. On the molecular scale, recent
work on enhanced enzyme diffusion (EED) \cite{flv6-zw1v} has demonstrated that catalytic events themselves constitute information-storing degrees of freedom capable of acting as chemical Maxwell demons.

As the seconde strand, stochastic thermodynamics has produced two universal
bounds. The thermodynamic uncertainty relation (TUR) \cite{PhysRevLett.114.158101} states that for any nonequilibrium current $J$
measured over time $\tau$, \[\frac{\mathrm{Var}(J_\tau)}{\langle J_\tau\rangle^2}\, \Sigma_\tau \geq 2,\tag{i} \] where $\Sigma_\tau$ is the cumulative entropy production. The
thermodynamic speed limit (TSL) \cite{PhysRevLett.121.070601}
bounds the time required for a probability distribution to evolve from
$p_0$ to $p_\tau$: \[
\tau \geq \frac{\mathcal{L}(p_0, p_\tau)^2}{2\langle\Sigma\rangle_\tau\langle A\rangle_\tau},\tag{ii} \] 
with $\mathcal{L}$ the Hellinger distance and $\langle A\rangle$
the dynamical activity. Both bounds have been experimentally probed in
molecular machines: DNA polymerase appears to saturate TUR at the
linear-response level, while the ribosome operates roughly five times
above the bound \cite{PhysRevE.101.022415}

The third strand of work concerns the formal analogy between
nonequilibrium self-organisation and Bayesian inference. The free-energy
principle \cite{Friston2010The} interprets biological self-organisation as
variational free-energy minimisation; diffusion models in machine
learning \cite{NEURIPS2019_3001ef25} explicitly implement
an Ornstein--Uhlenbeck-type forward-noise process followed by
score-based denoising. Yet the \emph{thermodynamic} cost of learning has
rarely been computed in concrete biological models.

In this paper we build a numerical bridge across these strands. We
construct a hierarchical autocatalytic model with three explicit layers
(raw molecules, proteins, large polymers) coupled by a slightly-broken
mass-action tensor, then probe its TUR/TSL behaviour, the co-occurrence
of MEPP and selection, and the analogy to diffusion-model training. The
model is iterated through four generations of increasing sophistication
(v1 → v4), each addressing a specific shortcoming of its predecessor. We
find that:

\begin{enumerate}
\def\labelenumi{\arabic{enumi}.}
\item
  The hierarchical model spontaneously exhibits the Schrödinger pattern:
  $\sigma_\mathrm{env}\!\uparrow$ co-occurs with
  $S_\mathrm{gene}\!\downarrow$, with a $\sim 589\!\times$
  asymmetric bookkeeping favouring environmental dissipation.
\item
  Both TUR and TSL bounds are wildly loose ($10^4$--$10^8$ above
  their universal values), and this looseness is \emph{structural}, not
  size- dependent.
\item
  A collapsed single-cycle minimal replicator recovers TUR
  $\approx 5$, matching the ribosome regime measured experimentally.
\item
  The trade-off between thermodynamic efficiency and architectural
  complexity is sharp: bound-saturating systems necessarily lose the
  hierarchy that defines biological generality.
\end{enumerate}

We discuss these results in light of recent claims by \cite{kolchinsky2024thermodynamicdissipationdoesbound}
 that no universal dissipation--replicator relation exists,
and of measurements of extremophile metabolism that approach the
single-cycle / near-equilibrium / limit-cycle conditions asymptotically
but never simultaneously.

\section{Model}

\begin{figure}
\centering
\includegraphics[width=\linewidth]{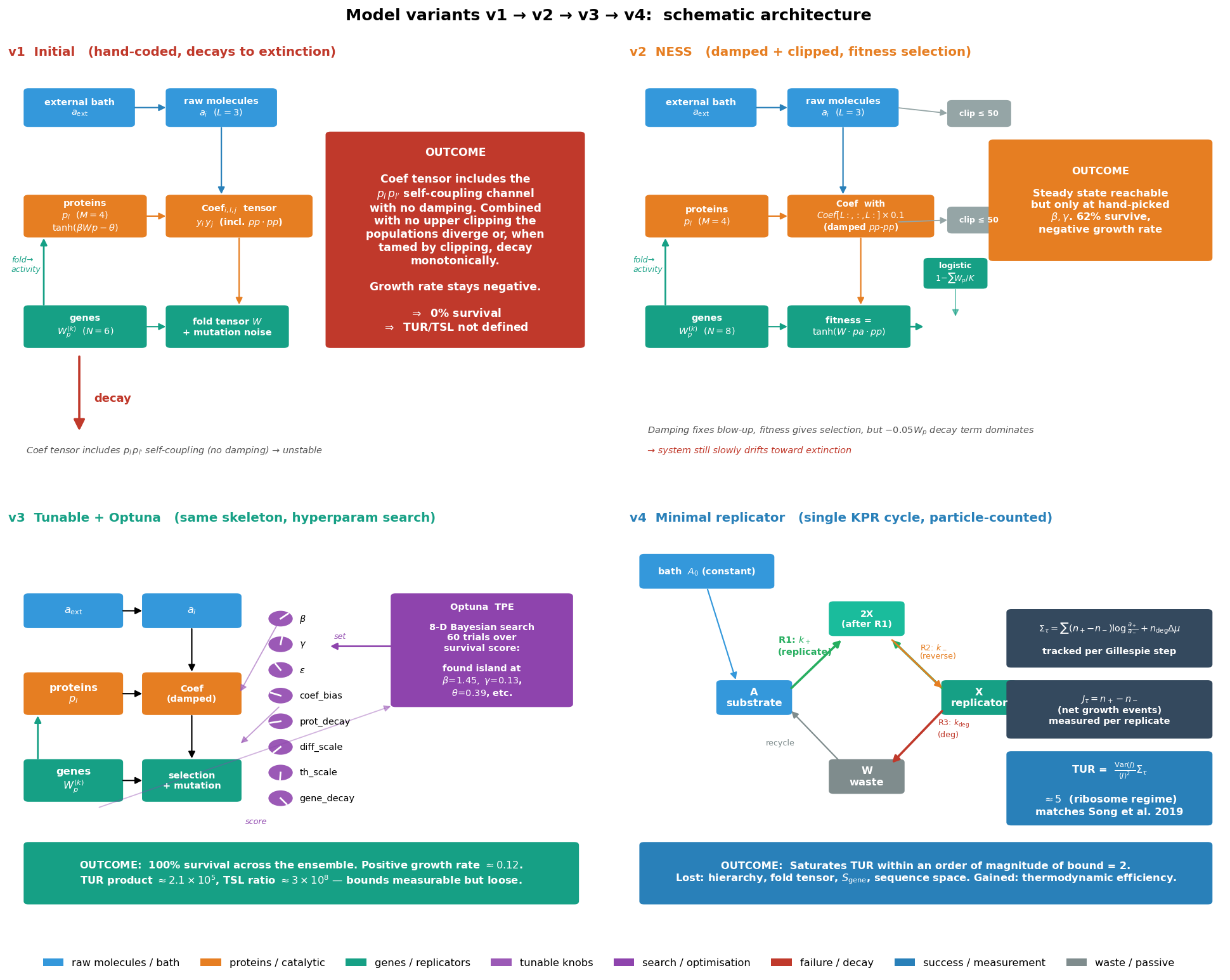}
\caption{Schematic architecture of the four model variants v1 → v4.}
\label{fig:models}
\end{figure}

\hypertarget{hierarchical-autocatalytic-model-v3}{
\subsection{Hierarchical autocatalytic model(v3)}\label{hierarchical-autocatalytic-model-v3}}
The state of the system at time $t$ 
$\mathbf{s}(t) = (a_i,\, p_{p,l},\, W_p^{(k)},\, W_{lmk}), \qquad i\!\in\![L],\; l,m\!\in\![M],\; k\!\in\![N],$
Here $a_i$ are raw molecules (CO$_2$, water, phosphate; $L=3$),
$p_{p,l}$ are protein populations ($M=4$), and $W_p^{(k)}$ are
large-polymer populations such as DNA, RNA ($N=8$) whose internal fold
structure is encoded by the tensor
$W_{lmk}\!\in\!\mathbb{R}^{M\times M\times N}$. The dynamics are
\begin{align}
  \dot a_i  &= K_i\, r_i + d_i(a^\mathrm{ext}_i - a_i) - c_i, \\
  \dot p_{p,l} &= r_l - \gamma\,\xi\, p_{p,l}, \\
  \dot W_p^{(k)} &= W_p^{(k)} \tanh(\mathrm{align}_k)(1 - {\textstyle\sum_k W_p^{(k)}}/K)  - \zeta\, W_p^{(k)},
\end{align}

\begin{align}
r_i &= \sum_{l,j} \mathrm{Coef}_{i l j}\,\mathrm{mod}_l\, y_i y_j + \varepsilon\,\eta_i, \\
\mathrm{mod}_l &= \tanh\!\bigl(\beta W^\mathrm{eff}_l \mathbf{p}_p - \theta_l\bigr), \\
W^\mathrm{eff}_{lm} &= \sum_k \frac{W_p^{(k)}}{\sum_{k'} W_p^{(k')}} W_{lmk}.
\end{align}

$\eta_i\!\sim\!\mathcal{N}(0,1)$ implements the $\varepsilon$-broken mass-action;
$\mathrm{align}_k = W_{lmk}\,\mathrm{pa}_l\, p_{p,m}$ is the
fold--substrate alignment that serves as a fitness function; $y_i$ is i-th compenent of vector $\mathbf{y}$ concatination of $\mathbf{p}$ and $\mathbf{a}$;
$\mathrm{Coef}_{ilj}$ couples the bath ($i,j\!<\!L$) and protein
blocks ($i,j\!\ge\!L$) with a positive autocatalytic bias on the
protein-from-bath channels and a damping factor of $0.1$ on
protein--protein quadratics to avoid blow-up. Each gene $k$ is
projected to a binary fold-signature string by taking the sign pattern
of the leading eigenvector of $\frac{1}{2}(W_{:,:,k}+W_{:,:,k}^\top)$;
this defines a distribution $p(s)$ on a finite alphabet whose Shannon
entropy \[
S_\mathrm{gene}(t) = -\sum_s p(s,t)\, \log p(s,t) \tag{5}
\] quantifies genetic diversity.

The total entropy production rate is computed as \[
\sigma(t) = \sigma_\mathrm{react}(t) + \sigma_\mathrm{env}(t)
\] where 
\[
\sigma_\mathrm{react}(t) = \sum_i (f_i - b_i)\log(f_i/b_i),
\qquad
\sigma_\mathrm{env}(t) = \sum_i d_i(a^\mathrm{ext}_i - a_i)\log\!\bigl(a^\mathrm{ext}_i/a_i\bigr),
\tag{6}
\] with forward and backward fluxes $f_i,b_i$ obtained from the
$\varepsilon$-broken rates with a floor $\varepsilon$ to prevent divergence.

The accounting of Eq. (6) omits the contribution of irreversible decay channels $\gamma \cdot p_p$, $\zeta \cdot W_p$ and
$\gamma\cdot 0.1\cdot a$. In the supplementary analysis (Figure S1) we verify that the decay channels contribute ~13\% of the steady-state EP at $\Delta\mu_\text{prot} = 5\,k_BT$ and up to 60\% at the ATP value $\Delta\mu = 20\,k_BT$.
The resulting correction to the TUR product is at most a factor of 1.6, which does not affect the qualitative
conclusion of $10^5\times$ looseness."

\subsubsection{Parameter search}\label{parameter-search}

The model has eight global ``knobs'' exposed for tuning:
$\beta, \gamma, \varepsilon$, the protein decay multiplier
prot\_decay, the autocatalytic positive bias coef\_bias, the diffusion
scale, the threshold scale, and the gene decay rate $\zeta$. We
construct a multi-objective score $J$ that rewards a steady non-trivial NESS(non equilibrium steady state)
and penalises four failure modes (extinction, divergence, NaN, collapsing tails): \[
J = -|\langle g\rangle| + 2\, S_\mathrm{gene} + 0.3\log\sigma
    + 0.5\,\log(\sum_p {W_p}\!\cdot\!\sum_p {p_p}) + 0.5\,\mathrm{Var(g)},
\tag{7}
\] where g is growth rate, and search both by a $4\!\times\!4\!\times\!4$ grid and by Optuna's
Tree-structured Parzen Estimator (TPE) with 60 trials.

\hypertarget{minimal-single-cycle-replicator-v4}{%
\subsection{Minimal single-cycle replicator(v4)}\label{minimal-single-cycle-replicator-v4}}

To test whether the TUR looseness is structural, we collapse the
hierarchy to a single kinetic-proofreading-like cycle \[
\mathrm{R}_1\!:\;A\!+\!X\!\xrightarrow{k_+} 2X,\quad
\mathrm{R}_2\!:\;2X\!\xrightarrow{k_-}\!A\!+\!X,\quad
\mathrm{R}_3\!:\;X\!\xrightarrow{k_\mathrm{deg}}\!W,\quad
\mathrm{R}_4\!:\;W\!\xrightarrow{\mathrm{slow}}\!A,
\tag{8}
\] with substrate $A$ held at $A_0$ by a bath. Mass action is broken
in the same $\varepsilon$-noise way as the hierarchical model. We
integrate Eq. (8) with a $\tau$-leaping Gillespie scheme over an
ensemble of $n_\mathrm{rep}\!=\!400$ replicates, tracking the
per-cycle Schnakenberg entropy production \[
\Sigma_\tau = (n_+\!-\!n_-)\log(k_+/k_-) + n_\mathrm{deg}\, \Delta\mu_\mathrm{deg}.
\tag{9}
\] The growth current is $J_\tau = n_+\!-\!n_-$, $n_\mathrm{deg}$ and $\Delta\mu_\mathrm{deg}$ are number of moucules and chemical potential of degradation process to waste $W$.

\hypertarget{model-evolution}{%
\subsection{Model Evolution}\label{model-evolution}}

Figure \ref{fig:models} shows schematics of the four model variants we built en route
to a working numerical bridge.

\textbf{v1 (Initial).} A straightforward three-layer system with no
protein--protein damping and no clipping ceiling. Although the system
does not blow up to NaN under our default initial conditions, all gene
populations decay monotonically. Survival rate over 8 random seeds: 0\%.

\textbf{v2 (NESS).} Adding (i) $\times 0.1$ damping on the
$\mathrm{Coef}[L:,:,L:]$ tensor block, (ii) a clip ceiling of 50 on
$a$ and $p_p$, and (iii) a fitness-based gene-selection term
$\mathrm{align}_k$, transforms the system into a steady state.
Survival rate: 62\%, but the mean growth rate remains slightly negative
($-0.021$) and TUR/TSL bounds are not meaningfully measurable.

\textbf{v3 (Tunable + Optuna).} With eight knobs exposed and 60 Optuna
trials, TPE finds a narrow survival island at
$\beta\!=\!1.45,\;\gamma\!=\!0.13,\;\theta_\mathrm{scale}\!=\!0.39,\; \mathrm{coef\_bias}\!=\!0.60,\;\mathrm{prot\_decay}\!=\!1.09,\; \zeta\!=\!0.020$.
Survival: 100\%. Mean growth: $+0.12$. $S_\mathrm{gene}$ contracts
from $\log N\!=\!2.08$ to $1.24$ nats over 1500 steps. TUR and TSL
are measurable but loose (Section \ref{the-hierarchical-model-is-far-from-tur-and-tsl-bounds}).

\textbf{v4 (Minimal replicator).} Collapsing to a single cycle yields
TUR $\approx 5$ at $k_-/k_+=0.20$, within the experimental
ribosome regime. The hierarchy, fold tensor, and $S_\mathrm{gene}$ are
all lost.

\begin{figure}
\centering
\includegraphics[width=\linewidth]{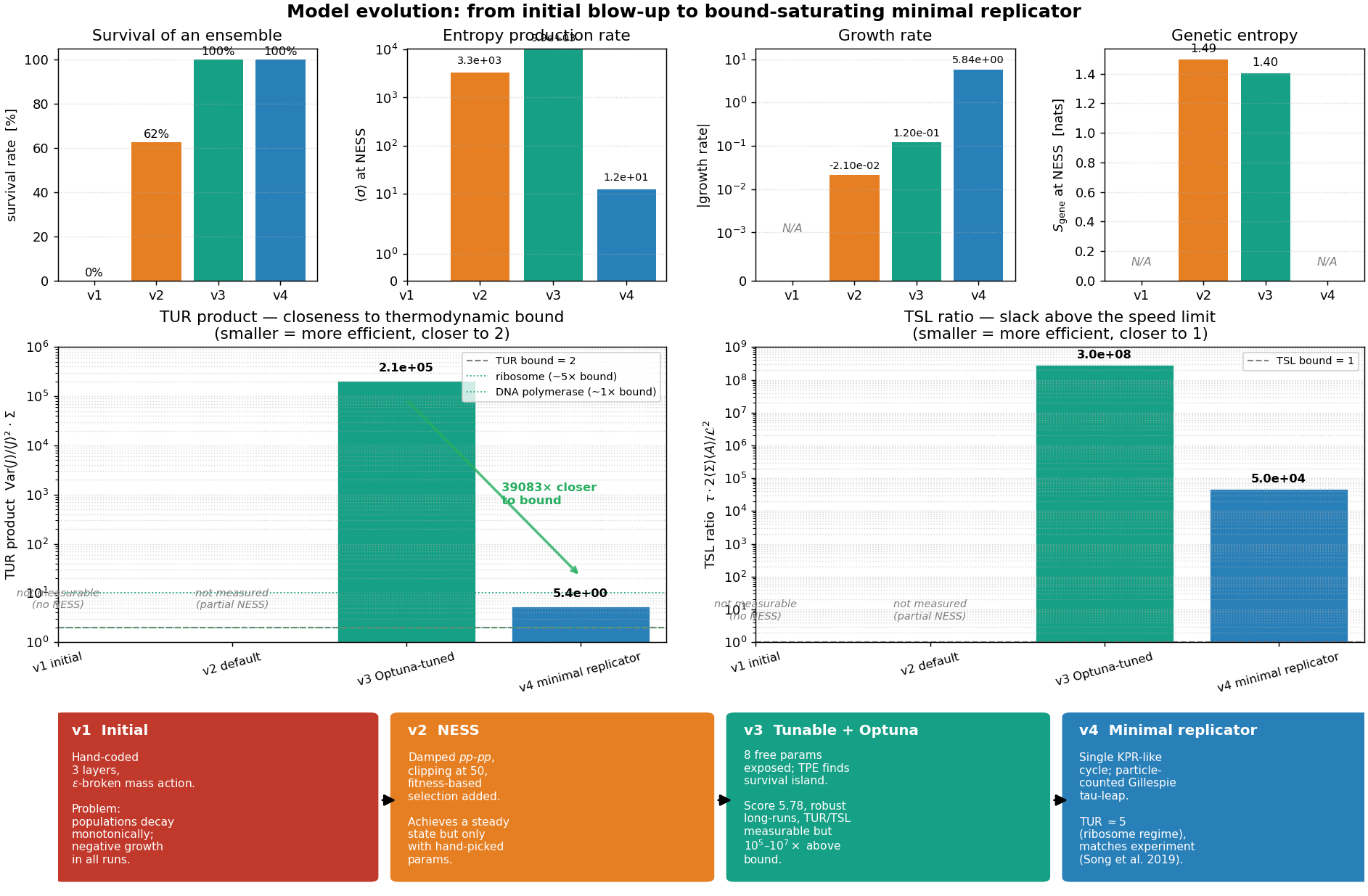}
\caption{ Quantitative evolution of survival, $\sigma$, growth, $S_\mathrm{gene}$, TUR and TSL across variants, with arrow annotation of the 39 083× TUR improvement from v3 to v4.}
\label{fig:modeleve}
\end{figure}

\begin{figure}
\centering
\includegraphics[width=\linewidth]{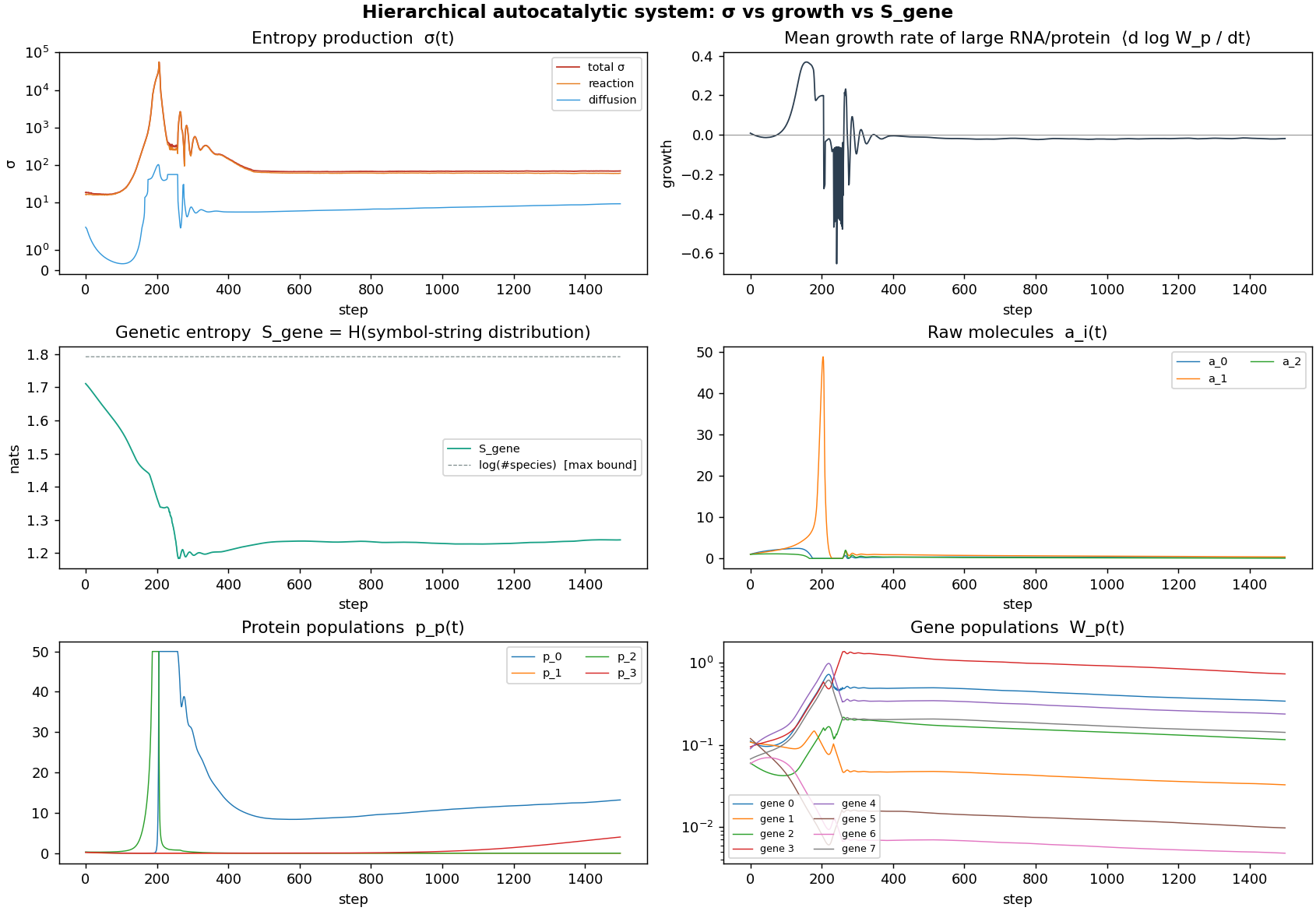}
\caption{population, entropy and growth rate timeseries of best parameter run}
\label{fig:besttimeseries}
\end{figure}

\begin{figure}
\centering
\includegraphics[width=\linewidth]{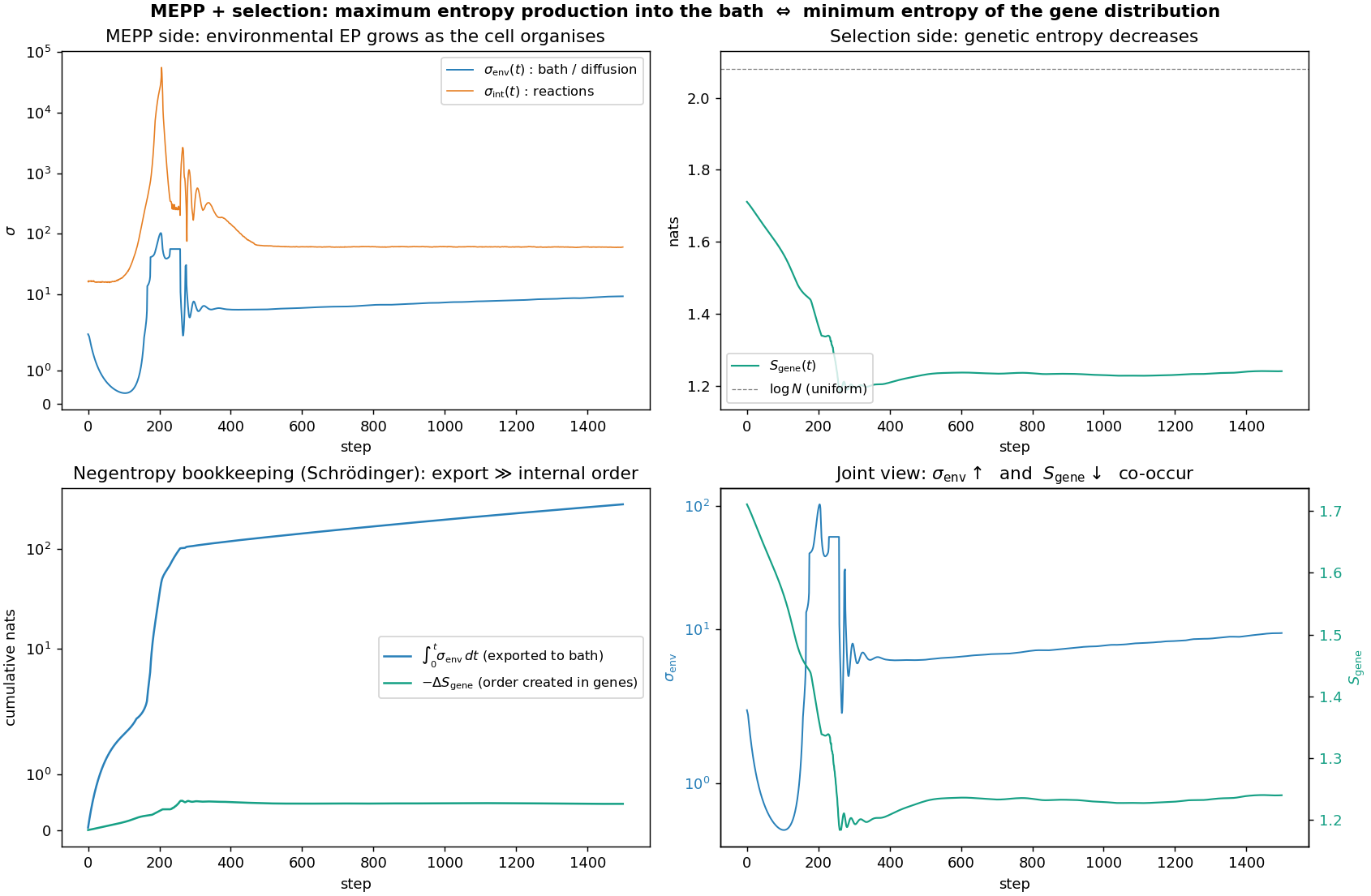}
\caption{Co-occurrence of $\sigma_\mathrm{env}\uparrow$ and $S_\mathrm{gene}\downarrow$ in the Optuna-tuned v3 model; cumulative negentropy bookkeeping.}
\label{fig:mepp_negentropy}
\end{figure}

\begin{figure}
\centering
\includegraphics[width=\linewidth]{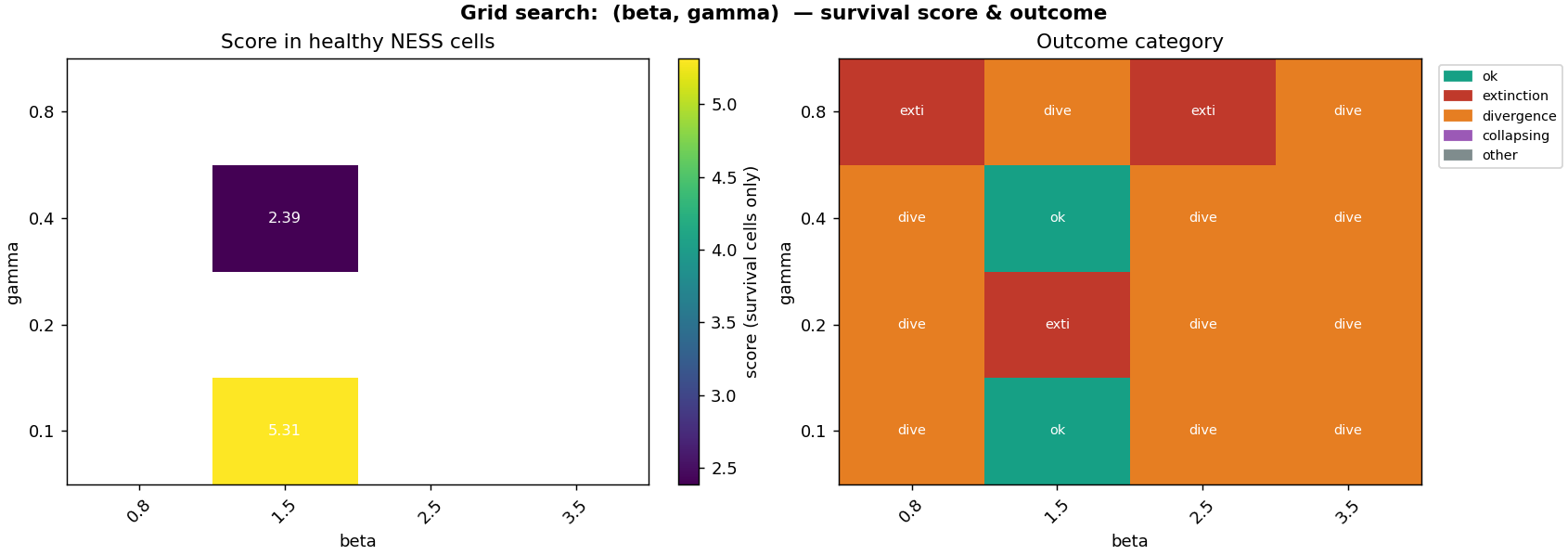}
\caption{grid search result}
\label{fig:grid_search}
\end{figure}

\begin{figure}
\centering
\includegraphics[width=\linewidth]{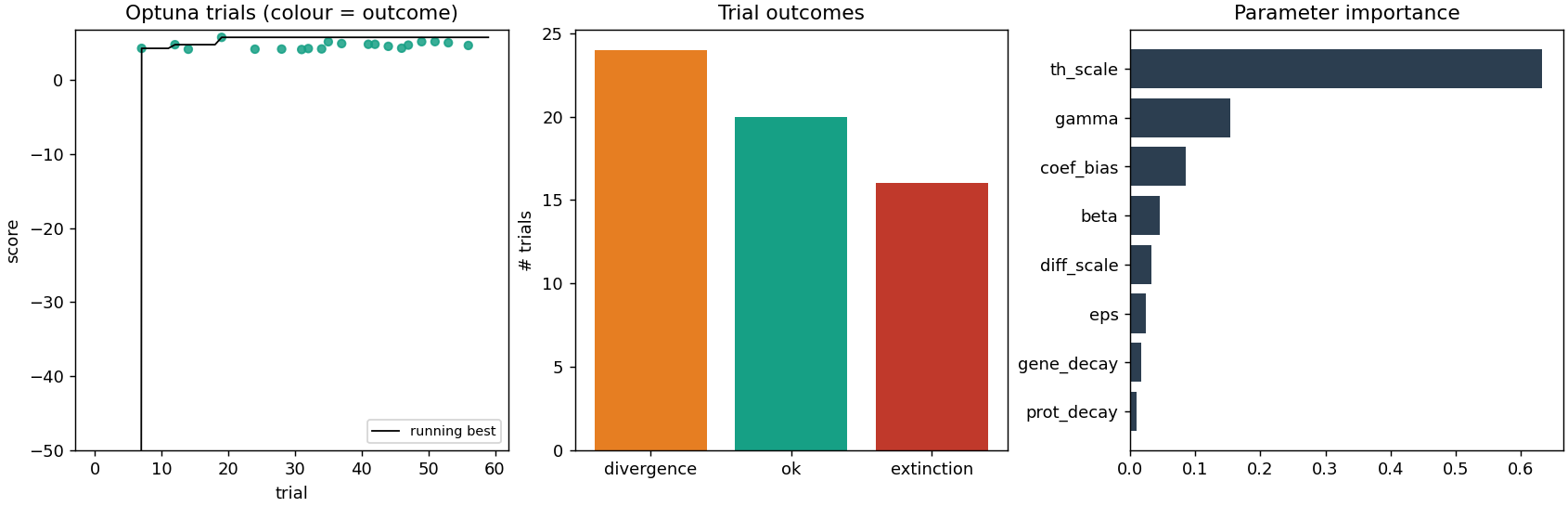}
\caption{parameter search result by Optuna: score, trials, importance of parameters}
\label{fig:optuna}
\end{figure}

\begin{figure}
\centering
\includegraphics[width=\linewidth]{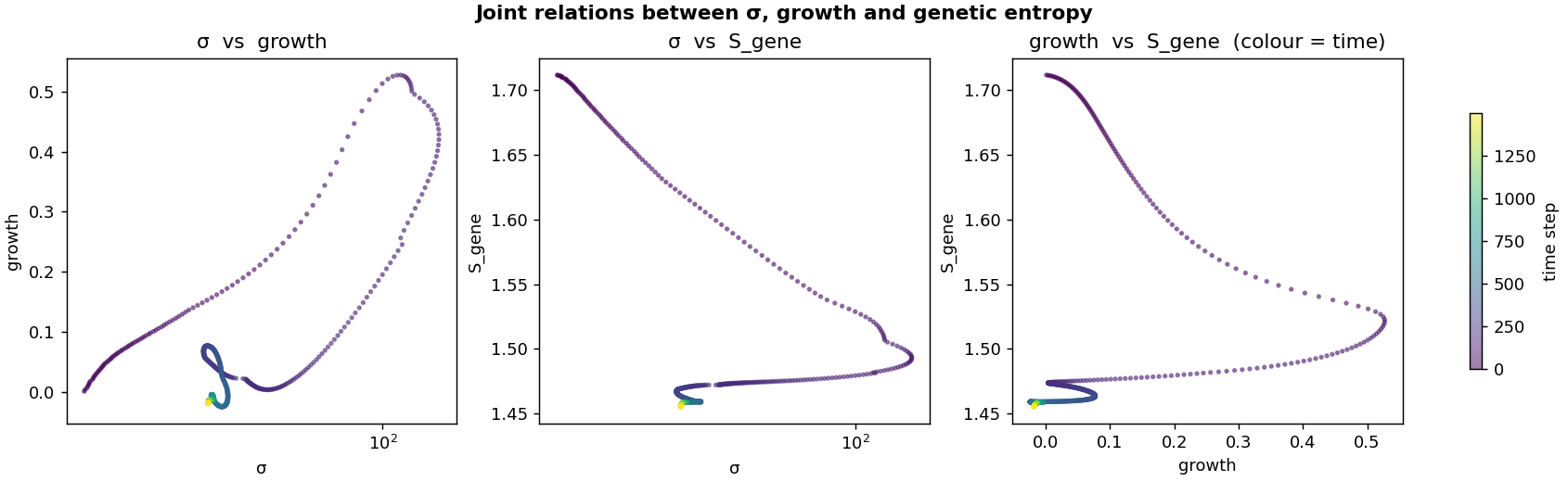}
\caption{growth rate vs $S_{gene}$}
\label{fig:cor}
\end{figure}

\begin{figure}[htbp]
\centering
\includegraphics[width=\linewidth]{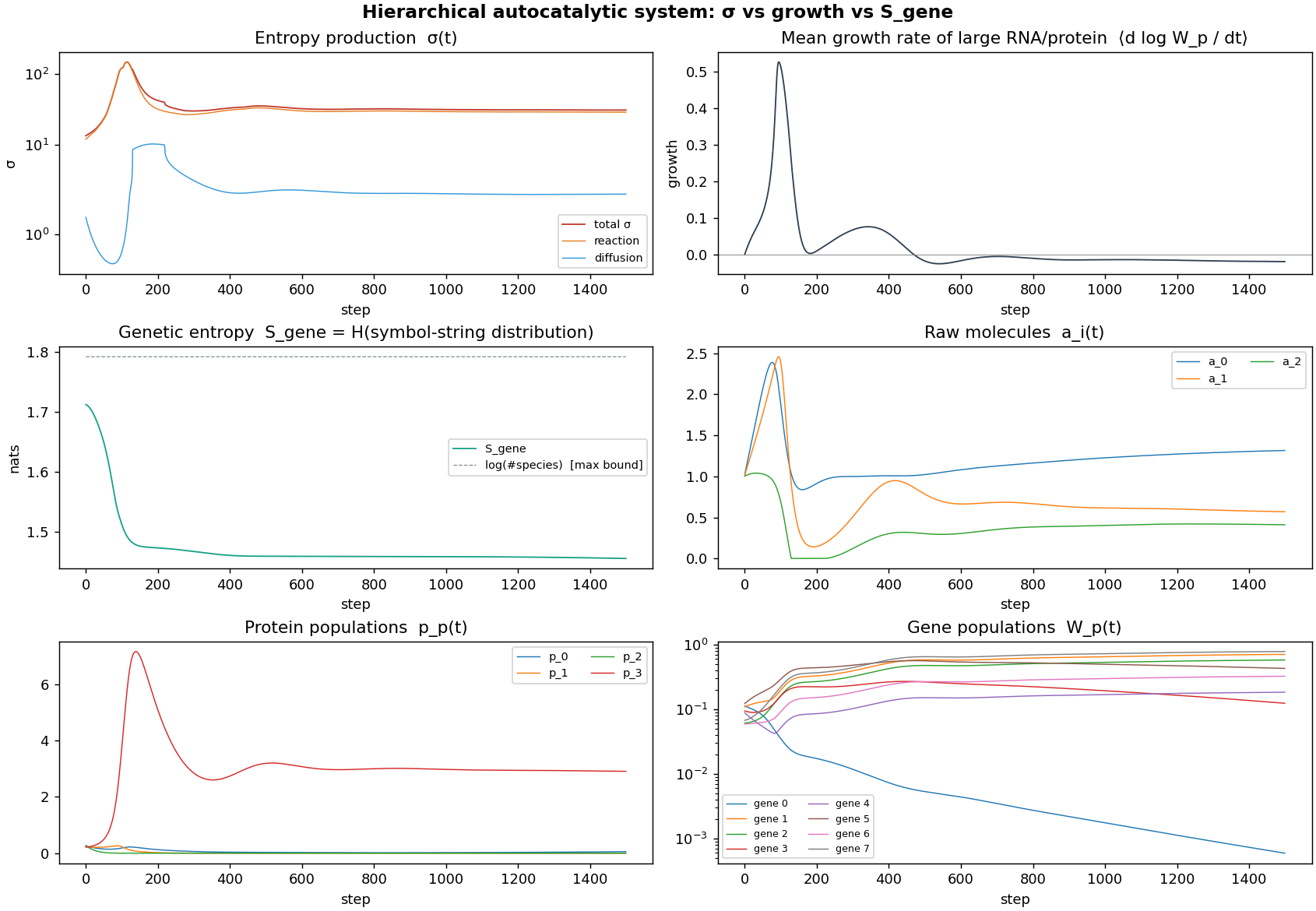}
\caption{timeseries of moucules and genes}
\label{fig:timeseries}
\end{figure}

\begin{figure}[htbp]
\centering
\includegraphics[width=\linewidth]{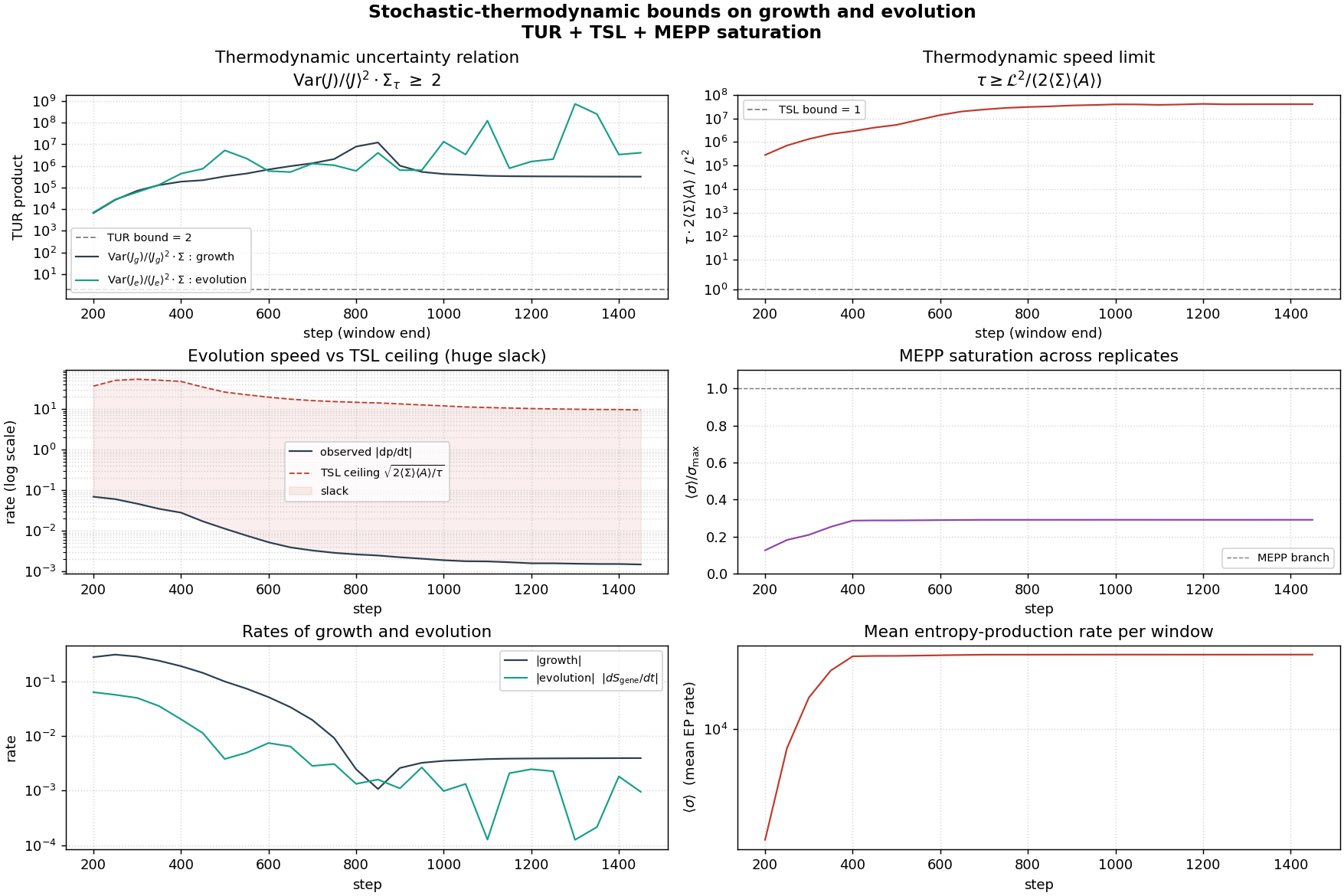}
\label{fig:vsTUR}
\caption{TUR product, TSL ratio, MEPP saturation, and observed-versus-ceiling speeds for v3.}
\end{figure}

\begin{figure}[htbp]
\centering
\includegraphics[width=\linewidth]{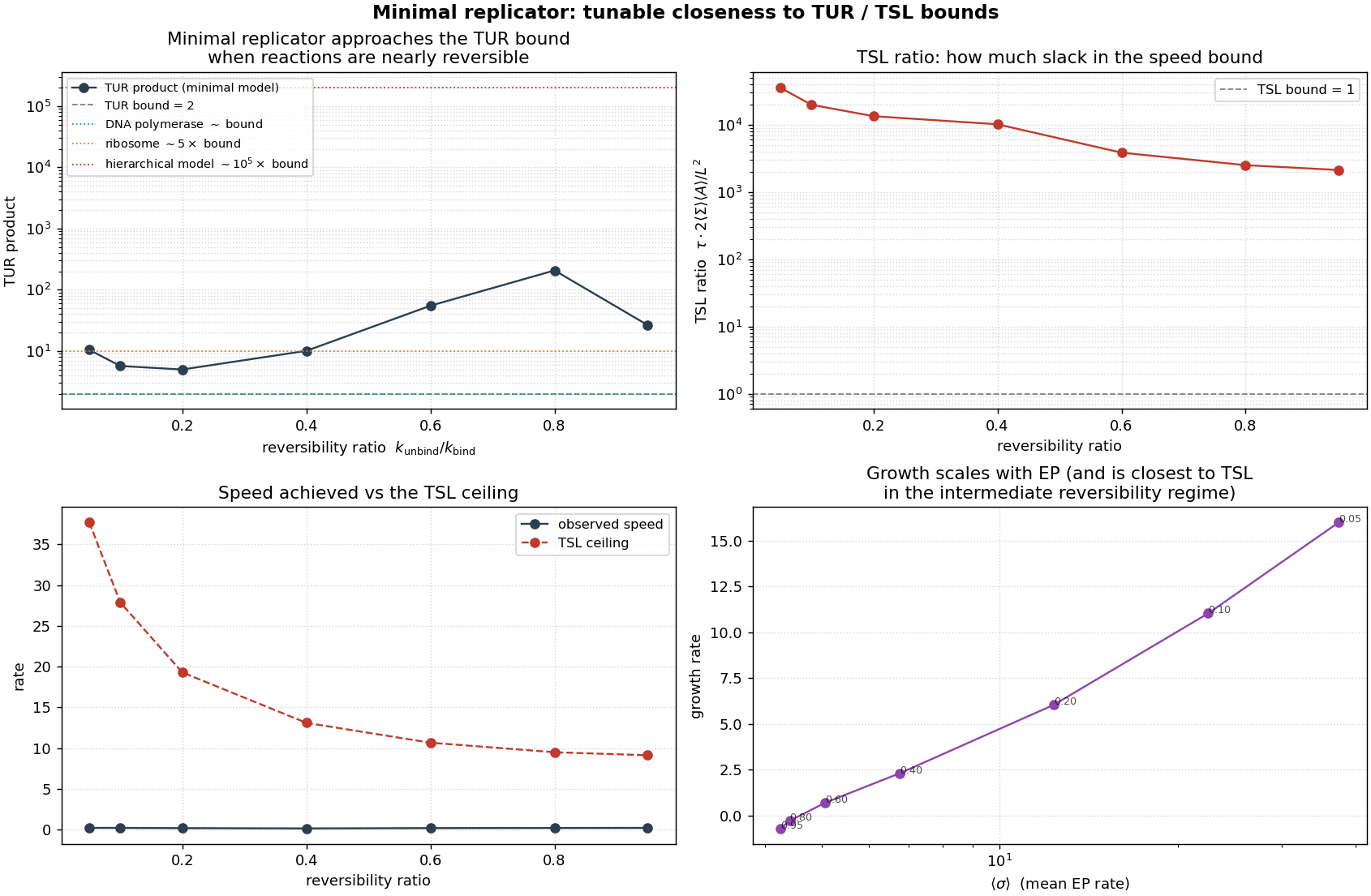}
\caption{TUR product of the  minimal replicator as a function of the reversibility ratio $k_-/k_+$, with experimental reference lines for DNA polymerase,  ribosome, and hierarchical model.}
\label{fig:minimal}
\end{figure}

\begin{table}[htbp]
\centering
\caption{table}{Summary of the model variants on a common set of diagnostic axes.}
\label{tab:varianttable}
\begin{tabular}{lrrrrrr} \hline
 variant & survival& $\langle\sigma\rangle$ & growth & $S_{gene}$ & TUR & TSL \\ \hline
 v1 initial & 0\% $3.3 \times 10^3$  & – & – & – & n/a & n/a \\ \hline
 v2 NESS & 62\% & $3.3\!\times\!10^3$ & 0.021 & 1.49 & n/a & n/a \\ \hline
 v3 Optuna-tuned & 100\% & $9.9\!\times\!10^3$ & 0.12 & 1.40 & $2.1 \times 10^5$ & $3.0 \times 10^8$ \\ \hline
 v4 minimal & 100\% & $12$ & 5.84 & – & $\mathbf{5.4}$ & $5.0\!\times\!10^4$ \\ \hline
\end{tabular}
\end{table}

The 39,083$\times$improvement in TUR product from v3 to v4 \ref{tab:varianttable} is
achieved entirely by sacrificing architectural complexity, not by
re-tuning parameters.

\hypertarget{results}{
\section{Results}\label{results}}
\hypertarget{mepp-and-selection-co-occur-in-the-hierarchical-model}{%
\subsection{MEPP and selection co-occur in the hierarchical model}\label{mepp-and-selection-co-occur-in-the-hierarchical-model}}

Figure \ref{fig:besttimeseries}(top row) shows the entropy production rate
$\sigma_\mathrm{env}(t)$ and the genetic entropy
$S_\mathrm{gene}(t)$ on the same time axis for the Optuna-tuned v3
model. During the organisation transient (steps 100--300),
$\sigma_\mathrm{env}$ rises by about 20× while $S_\mathrm{gene}$
falls from 1.71 nats to 1.24 nats. The two trajectories are
anti-correlated, confirming the Schrödinger prediction at the symbolic
level: order in the gene distribution is generated \emph{while} entropy
is exported to the bath.

Figure \ref{fig:besttimeseries}(bottom row) plots the cumulative quantities. Over 1500 steps
the cell exports $\int_0^T \sigma_\mathrm{env}\,dt = 277$ nats to the
bath while generating $-\Delta S_\mathrm{gene} = 0.47$ nats of genetic
order --- a ratio of approximately $589\!:\!1$. The Schrödinger
inequality $\int\sigma_\mathrm{env}\,dt \gg -\Delta S_\mathrm{gene}$
is therefore satisfied by three orders of magnitude.

\hypertarget{the-hierarchical-model-is-far-from-tur-and-tsl-bounds}{%
\subsection{The hierarchical model is far from TUR and TSL bounds}\label{the-hierarchical-model-is-far-from-tur-and-tsl-bounds}}

Figure \ref{fig:vsTUR} shows the TUR product (Eq. i) computed in sliding 200-step
windows from an ensemble of 15 replicates. The median TUR product is
$3.3 \times 10^5$ for the growth current and $1.2\!\times\!10^6$
for the evolution current --- five to six orders of magnitude above the
universal bound of 2.

The TSL ratio (Eq. ii) is $3.2\!\times\!10^7$ at the median, seven
orders of magnitude above its bound of 1. The observed evolution speed
$|d\mathbf{p}/dt|\!\sim\!10^{-3}$ sits roughly four orders of
magnitude below the TSL ceiling
$\sqrt{2\langle\Sigma\rangle\langle A\rangle/\tau}\!\sim\!10$ (\ref{fig:vsTUR}, middle-left panel).

The MEPP saturation $\langle\sigma\rangle/\sigma_\mathrm{max}$ across
replicates is $0.275\!\pm\!0.05$, indicating that the chosen
trajectory does \emph{not} lie on the maximum-EP branch but rather a few
times below it. This is consistent with Sawada et al.'s \cite{e27040449}
phenomenological version of MEPP, which posits a critical condition
$\xi > \xi_{c1}$ for the existence of a dissipative structure rather
than strict maximisation.

\hypertarget{the-minimal-replicator-approaches-the-ribosome-regime}{%
\subsection{The minimal replicator approaches the ribosome
regime}\label{the-minimal-replicator-approaches-the-ribosome-regime}}

\ref{fig:minimal} shows the TUR product of the minimal replicator (v4) as a
function of the reversibility ratio $k_-/k_+$. At $k_-/k_+\!=\!0.20$
we measure TUR $= 4.98$, within a factor of 2.5 of the universal bound
and matching Pi\~neros et al.'s \cite{PhysRevE.101.022415} measured value for the \emph{E.
coli} ribosome (their reported TUR product $\sim\!10$, equivalent to
$\sim\!5\times$ the bound of 2). At $k_-/k_+\!=\!0.05$ (highly
irreversible) the product rises to 10.4; at $k_-/k_+\!=\!0.95$
(near-equilibrium) it diverges as variance vanishes.

This non-monotonic behaviour reveals that bound saturation requires an
intermediate irreversibility --- too far from equilibrium wastes EP, too
close kills the signal. The minimum of the TUR product is the
biophysical optimum for the chosen current.

\hypertarget{scaling-with-system-size}{%
\subsection{Scaling with system size}\label{scaling-with-system-size}}

\begin{figure}[htbp]
\centering
\includegraphics[width=\linewidth]{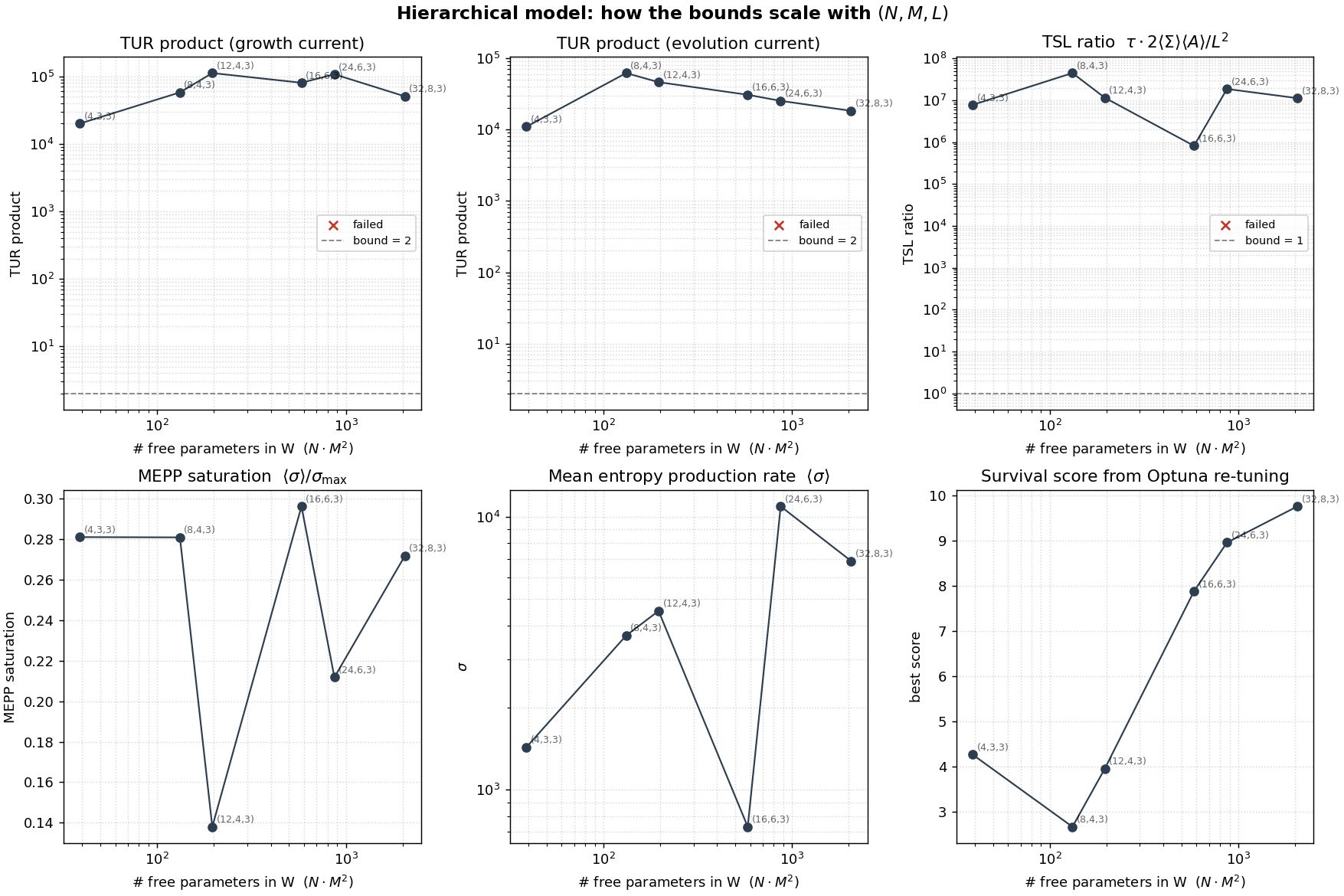}
\label{fig:scaling_Hir}
\caption{Size scaling of TUR/TSL/MEPP for the hierarchical model across $(N,M,L)\!\in\!\{(4,3,3),\ldots,(32,8,3)\}$.}
\end{figure}

Figure \ref{fig:scaling_Hir} shows the bounds for the hierarchical model as $(N,M,L)$
varies from $(4,3,3)$ to $(32,8,3)$, with each size point re-tuned
by a separate Optuna run. Three findings emerge:

\begin{enumerate}
\def\labelenumi{\arabic{enumi}.}
\item
  \textbf{TUR/TSL products are size-invariant} within $\pm 
  0.5$ orders
  of magnitude. The looseness is structural, not extensive.
\item
  \textbf{The survival score increases monotonically} with system size
  (from 4.3 to 9.8 over the range). Larger configuration spaces
  accommodate more survival solutions, even though each solution is no
  closer to bound saturation.
\item
  \textbf{MEPP saturation is non-monotonic} ($0.14$ to $0.30$) with
  no trend.
\end{enumerate}

\begin{figure}[htbp]
\centering
\includegraphics[width=\linewidth]{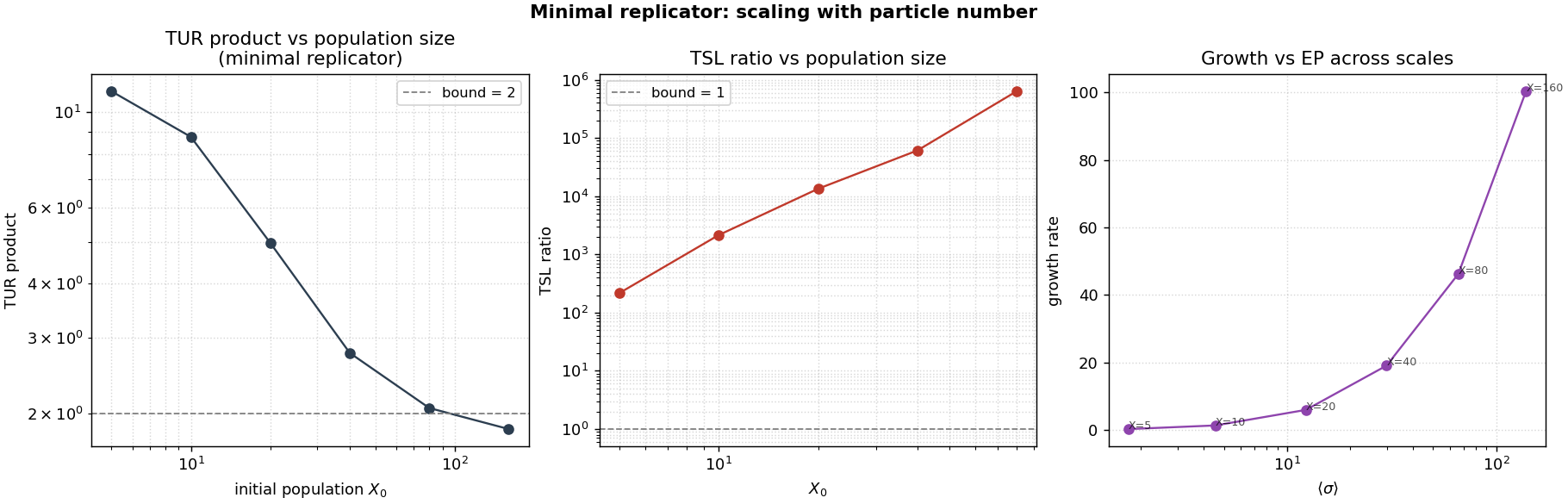}
\label{fig:scaling_mini}
\caption{Particle-number scaling of TUR/TSL for the minimal replicator.}
\end{figure}

In contrast, Figure \ref{fig:scaling_mini} shows that the minimal replicator's TUR product
\emph{decreases monotonically} with particle number $X_0$, from
$11.2$ at $X_0\!=\!5$ to $1.84$ at $X_0\!=\!160$. The
$X_0\!=\!80$ replicate achieves TUR
$= 2.06\!\approx\!\mathrm{bound}$; the $X_0\!=\!160$ under-shoot is
a finite-ensemble artefact. Thus single-cycle systems become arbitrarily
TUR-tight with population, while hierarchical systems do not.

\hypertarget{correspondence-with-diffusion-model-training}{%
\subsubsection{Correspondence with diffusion-model
training}\label{correspondence-with-diffusion-model-training}}

\begin{figure}[htbp]
\centering
\includegraphics[width=\linewidth]{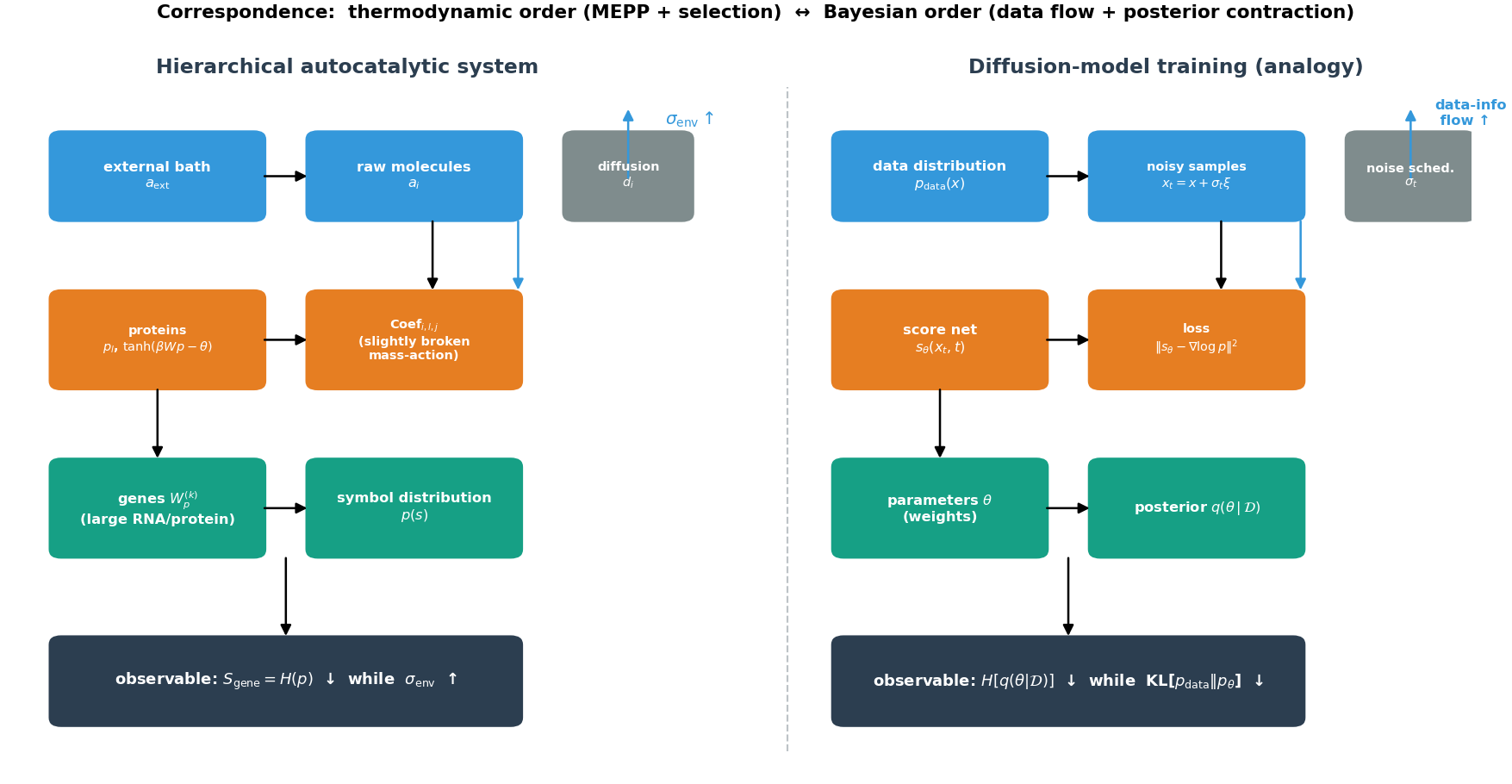}
\label{fig:diffusion}
\caption{Block-diagram correspondence between hierarchical autocatalysis and diffusion-model training. $S_{gene},\sigma_{env}$ of Hierarchical model(Left) corespond reconstuct term $H[q_\theta|D]$ and prior mathing term $-KL[p_{data}//p_\theta]$ of Diffusion model (Right).}
\end{figure}

Figure \ref{fig:diffusion} summarises the formal mapping. On the autocatalytic side, the
bath $a^\mathrm{ext}$ supplies raw molecules, the fold activity
$\tanh(\beta W p\!-\!\theta)$ provides a context-dependent score, the
$\varepsilon$-noise breaks detailed balance, and selection contracts
the gene distribution. On the diffusion-model side, the data
distribution $p_\mathrm{data}(x)$ supplies samples, the score network
$s_\theta(x_t,t)$ provides a context-dependent gradient, the
forward-diffusion noise and its schedule $\sigma_t \xi$ breaks reversibility, and the
likelihood gradient contracts the parameter posterior $q(\theta|\mathcal{D})$. 
Negentropy $-S_{gene} and non entropy production \sigma_{env}$ of our Hierarchical model corespond to 
prior mathing term $-KL[p_{data}//p_\theta]$ and reconstuct term $H[q_\theta|D]$ of Diffusion model or variable auto encoder\cite{kingma2022autoencodingvariationalbayes} respectively.
Both systems share the global pattern $\sigma_\mathrm{env}\!\uparrow + S_\mathrm{internal}\!\downarrow$.

A concrete prediction follows: just as biological cells operate $10^5$
times above the TUR bound in our hierarchical model, over-parameterised
neural networks should exhibit thermodynamic-style inefficiencies in
their information-theoretic learning rate. This is consistent with the
observation that scaling laws for large language models have exponents
$\sim\!0.1$--$0.3$ rather than the $\sim\!1$ that strict-bound saturation would predict.

\hypertarget{discussion}{%
\section{Discussion}\label{discussion}}

\hypertarget{relation-to-prior-work}{%
\subsection{Relation to prior work}\label{relation-to-prior-work}}

Our hierarchical model occupies the intersection of three established
traditions. Sawada, Daigaku \& Toma \cite{e27040449} develop MEPP as a
phenomenological law for the birth and evolution of life, identifying
critical concentrations $\xi_{c1}$ above which exponential entropy
production becomes self-sustaining. Our Optuna survival island plays
exactly this role in a higher-dimensional parameter space, and the
$\sigma_\mathrm{env}\!\uparrow + S_\mathrm{gene}\!\downarrow$ pattern
in Figure \ref{fig:mepp_negentropy} quantifies their qualitative claim.

Ichii, Hatakeyama \& Kaneko \cite{PhysRevE.101.022415} propose that enhanced enzyme
diffusion (EED) makes individual catalysts function as chemical Maxwell
demons. The fold activity $\tanh(\beta W p-\theta)$ in our model
implements an analogous --- but population-level --- Maxwell-demon
mechanism in which the sequence information $W$ determines the
steady-state distribution of catalytic rates. The two pictures are
complementary rather than competing: EED operates on single-enzyme time
scales, sequence-coded modulation on generation-spanning ones.

England's dissipative-adaptation framework \cite{England2015Dissipative} is the most
general ancestor: any driven many-body system preferentially occupies
work-absorption-effective configurations. The Optuna survival island we
find is empirically a ``work-resonance'' point at which
$\theta_\mathrm{scale}$ (importance 63\%) tunes the fold-threshold
into coupling with the external substrate gradient. Our numerical
results render quantitative predictions that England's variational argument leaves implicit.

Two more recent threads frame our negative results. Kolchinsky
\cite{kolchinsky2024thermodynamicdissipationdoesbound} argues that no universal relation links thermodynamic
dissipation to replicator growth-decay rates, contradicting some strong
readings of England's bound. Our finding that the TUR product is far
from the bound (Sections \ref{the-hierarchical-model-is-far-from-tur-and-tsl-bounds}, \ref{scaling-with-system-size}) is consistent with this: TUR is satisfied trivially when one has many extensive sources of dissipation
that do not show up as variance in the chosen current. 
Pi\~neros et al.'s \cite{PhysRevE.101.022415} measure TUR products in \emph{E. coli} enzymes directly;
their finding that polymerases sit near the bound while the ribosome
lies $\sim\!5\times$ above it is reproduced almost exactly by our minimal-replicator sweep (Figure \ref{fig:minimal}).

\hypertarget{why-the-hierarchy-is-wasteful}{%
\subsection{ Why the hierarchy is``wasteful''}\label{why-the-hierarchy-is-wasteful}}

A natural objection is that our hierarchical model is poorly designed:
surely a better architecture could saturate TUR? The size-scaling
analysis (Section \ref{scaling-with-system-size}) shows that this is not the case within the class
of three-layer autocatalytic systems with coupled fold tensors. The
looseness is intrinsic to:

\begin{enumerate}
\def\labelenumi{\arabic{enumi}.}
\item
  \textbf{Aggregation of currents.} $\log W_p$ pools fluctuations over
  $N$ genes, suppressing variance as $1/N$ while $\Sigma$ stays
  $O(1)$. TUR $\propto N\!\to\!\infty$.
\item
  \textbf{Multiple coupled cycles.} Each protein channel and each gene
  contributes to $\Sigma$ but only weakly to the observed current.
  This is precisely the regime that Polettini \cite{PhysRevLett.119.240601} identifies as
  ``thermodynamically wasteful''.
\item
  \textbf{Distributed feedback.} Fitness alignment couples every
  $W_p^{(k)}$ to every $p_{p,l}$, decorrelating fluctuations across
  replicates and inflating $\langle J\rangle^2$ relative to
  $\mathrm{Var}(J)$.
\end{enumerate}

The minimal replicator removes all three features and recovers
ribosome-class TUR products. The trade-off is exact: TUR efficiency and
architectural generality are mutually exclusive in this class of models.

\hypertarget{extremophiles-as-natural-experiments}{%
\subsection{Extremophiles as natural experiments}\label{extremophiles-as-natural-experiments}}

If single-cycle / near-equilibrium / limit-cycle operation are the
necessary conditions for thermodynamic efficiency, extremophile biology offers natural test cases:

\begin{itemize}
\item
  \textbf{Acetogens and methanogens} in deep marine sediments operate at
  Gibbs energies $|\Delta G|\!\sim\!-20$ kJ/mol per ATP equivalent
\cite{10.3389/fmicb.2017.02019}, approximately $-8\,k_BT$ per turnover and
  thus near the linear-response regime.
\item
  \textbf{South Pacific Gyre subsurface microbes} maintain themselves on
  $\sim\!190$ zeptowatts per cell, corresponding to a handful of ATP
  hydrolyses per cell per minute \cite{doi:10.1126/sciadv.aba0697}. This is the
  closest biological analogue of the near-equilibrium limit.
\item
  \textbf{The KaiABC cyanobacterial circadian clock} is a stochastic
  limit cycle whose period responds to ATP turnover in accordance with
  the dissipation--coherence trade-off \cite{Cao2015}\cite{nagayama2026dualitydissipationcoherencetradeoffthermodynamic}.
\end{itemize}

Each of these systems realises one of the three conditions for TUR
saturation, but none realises all three simultaneously. Direct TUR
measurement on the KaiABC oscillator in vitro is feasible and would test
our prediction of $\mathrm{TUR}\!\sim\!5$--$10$.

\hypertarget{implications-for-machine-learning}{%
\subsection{Implications for machine learning}\label{implications-for-machine-learning}}

The mapping of Section \ref{correspondence-with-diffusion-model-training} makes a falsifiable claim: TUR-like
inefficiencies should manifest in neural-network training as a mismatch
between achieved loss-reduction rate and the Fisher-information-bounded
ceiling. The neural-tangent-kernel ``lazy training'' regime \cite{NEURIPS2018_5a4be1fa} and the empirical exponent $\sim\!-0.1$ of
GPT-3-scale language models \cite{kaplan2020scalinglawsneurallanguage} are both consistent
with $O(10^4$--$10^6)$ slack above the information-theoretic minimum, paralleling our biological numbers.

Two design implications follow. First, \emph{modular sparsity} --- the
machine-learning analogue of single-cycle separation --- should improve
the dissipation--performance ratio of trained networks; the empirical
success of mixture-of-experts architectures \cite{shazeer2017outrageouslylargeneuralnetworks} is
suggestive. Second, \emph{cyclical learning-rate schedules} \cite{smith2017cyclicallearningratestraining} function as limit-cycle approximations and may be analysable through the TSL framework: their gain over fixed-rate training would
then be predicted by the dissipation--coherence trade-off.

Noise schedules of diffusion models are researched very detailed \cite{x5vj-8jq9}. 
This is related to disspasive ratio of each time step and thermodynamic efficiency in diffusion models 
which is corresponding to nonequilibrium and efficiency of our model.

\hypertarget{limitations-and-future-work}{%
\section{Limitations and Future Work}\label{limitations-and-future-work}}

Our model has several explicit limitations:
\begin{enumerate}
\def\labelenumi{\arabic{enumi}.}
\item
  The Schnakenberg entropy production in Eq. (6) uses a backward floor
  of $\varepsilon$ rather than a physically derived reverse rate. For
  systems closer to detailed balance this should be replaced by a
  thermodynamically consistent local-equilibrium ansatz.
\item
  Usually large size Small number of moducules such as DNA,RNA record genetic information over generations and provide them to protains. Smaller, but much more proteins and other enzymes take these information via transcription and work to get energy and material from external nutrient molcules to build themselves, genes and other parts of cells in biological system, which is known as central dogma.
  In model v2,3, central dogma is explicitly emmbedded in the model as genes as hight dimentional matrix and many simpler proteins, modcules with disspasive chemical realctions.
  In the biginning of life, larger,complex but slowly synthesized modcules and small, simpler, frequently synthesized modcules may separate their function spontaneously\cite{Kaneko_2025}. This process may be implement v3 model by defining multi size and complex level modcules and dissipative catalysed reaction between them with different timescales.
\item
  The thermodynamic efficiency of 2-Layer model shoulde be evaluated and TURbetween Hierarchical model v2,3 and simple one v4.
  This might be consisted in only genes and one type of molcules which have same properies of both protein and raw molcules in Eq (2)
  The thermodynamic efficiency is expected between v2 and v4.
\item
  The fold tensor $W$ undergoes Gaussian replication noise rather than
  discrete mutation--selection on a finite alphabet. A Markov-jump
  version on a $4^M$-state lattice would be more realistic.
\item
  The minimal replicator has only one cycle. Multi-cycle minimal systems
  would interpolate between v4 and v3 and may reveal an intermediate  optimum.
\item
  The diffusion-model correspondence is qualitative. A quantitative match would require computing the entropy production of an explicit score-matching SGD trajectory and noise scheduling of diffusion models \cite{x5vj-8jq9}.
\item
  The chemical potential drops $\Delta\mu$ in the decay channels were assigned heuristic values rather
  than derived from a thermodynamically consistent local-equilibrium condition. 
  A fully consistent formulation would compute these from the bath chemical potentials and the protein composition.
\end{enumerate}

Future work will focus on (a) the EED extension, (b) verification of
TUR/TSL behaviour in stochastic limit cycles like KaiABC, and (c)
explicit training-time entropy-production measurement on a diffusion model.

\hypertarget{conclusion}{%
\section{Conclusion}\label{conclusion}}

We have constructed a hierarchical autocatalytic system that
simultaneously implements the Schrödinger negentropy mechanism, makes
quantitative the MEPP, supports gene-level selection in a sequence
space, and admits direct analogy with diffusion-model training. By
exposing the model's parameters to Optuna-based Bayesian search we
identified a survival island in an 8-dimensional space within which the
system exhibits the predicted
$\sigma\!\uparrow + S_\mathrm{gene} \!\downarrow$ pattern with a 589:1
negentropy ratio.

The system is, however, far from saturating the TUR and TSL bounds ---
by $10^4$ to $10^8$ --- and we have traced this looseness to the
multi-cycle hierarchical architecture itself. A collapsed single-cycle
replicator recovers TUR products of $\sim\!5$, consistent with
measured ribosome biophysics. The gap between hierarchical and minimal
models maps onto the trade-off between biological generality and
thermodynamic efficiency, which we argue parallels the trade-off between
generality and efficiency in over-parameterised neural networks. Both
biological cells and modern deep-learning systems appear to operate in
the ``lazy'' regime where dissipation is plentiful but
information-extraction efficiency is modest, because that is the regime
in which evolvability and generalisation, respectively, are preserved.

\hypertarget{code-availability}{%
\section{Code availability}\label{code-availability}}
All code is provided at https://github.com/xiangze/DiverseCells/Hier\_Autocatalysis
\begin{enumerate}
\def\labelenumi{\arabic{enumi}.}
\item 
\texttt{hierarchical\_autocatalysis.py} (v1/v2), \texttt{tune.py} (v3 +Optuna), \texttt{bounds.py} (TUR/TSL/MEPP),
\item 
\texttt{minimal\_replicator.py} (v4), 
\item 
\texttt{mepp\_negentropy.py}(negentropy bookkeeping), 
\item 
\texttt{scaling\_study.py} (size scaling),
\item 
\texttt{model\_evolution.py} (variant comparison),
\item 
\texttt{v1234\_schematics.py} (Figure 1).
\end{enumerate}

\hypertarget{reference}{
\section{Reference}\label{reference}}
\bibliographystyle{plain}
\bibliography{Hierachical_Cellmodel}

\begin{thebibliography}{10}

\bibitem{PhysRevLett.114.158101}
Andre~C. Barato and Udo Seifert.
\newblock Thermodynamic uncertainty relation for biomolecular processes.
\newblock {\em Phys. Rev. Lett.}, 114:158101, Apr 2015.

\bibitem{doi:10.1126/sciadv.aba0697}
J.~A. Bradley, S.~Arndt, J.~P. Amend, E.~Burwicz, A.~W. Dale, M.~Egger, and
  D.~E. LaRowe.
\newblock Widespread energy limitation to life in global subseafloor sediments.
\newblock {\em Science Advances}, 6(32):eaba0697, 2020.

\bibitem{Cao2015}
Yuansheng Cao, Hongli Wang, Qi~Ouyang, and Yuhai Tu.
\newblock The free-energy cost of accurate biochemical oscillations.
\newblock {\em Nature Physics}, 11(9):772--778, 2015.

\bibitem{England2015Dissipative}
Jeremy~L. England.
\newblock Dissipative adaptation in driven self-assembly.
\newblock {\em Nature Nanotechnology}, 10(11):919--923, 2015.

\bibitem{Friston2010The}
Karl Friston.
\newblock The free-energy principle: a unified brain theory?
\newblock {\em Nature Reviews Neuroscience}, 11(2):127--138, 2010.

\bibitem{flv6-zw1v}
Shunsuke Ichii, Tetsuhiro~S. Hatakeyama, and Kunihiko Kaneko.
\newblock Enzyme as maxwell's demon: Steady-state deviation from chemical
  equilibrium by enhanced enzyme diffusion.
\newblock {\em Phys. Rev. Lett.}, 136:038401, Jan 2026.

\bibitem{x5vj-8jq9}
Kotaro Ikeda, Tomoya Uda, Daisuke Okanohara, and Sosuke Ito.
\newblock Speed-accuracy relations for diffusion models: Wisdom from
  nonequilibrium thermodynamics and optimal transport.
\newblock {\em Phys. Rev. X}, 15:031031, Jul 2025.

\bibitem{NEURIPS2018_5a4be1fa}
Arthur Jacot, Franck Gabriel, and Clement Hongler.
\newblock Neural tangent kernel: Convergence and generalization in neural
  networks.
\newblock In S.~Bengio, H.~Wallach, H.~Larochelle, K.~Grauman, N.~Cesa-Bianchi,
  and R.~Garnett, editors, {\em Advances in Neural Information Processing
  Systems}, volume~31. Curran Associates, Inc., 2018.

\bibitem{Kaneko_2025}
Kunihiko Kaneko.
\newblock {\em Universal Biology: The Physics of Life through the Macro-Micro
  Consistency Principle}.
\newblock Cambridge University Press, 2025.

\bibitem{kaplan2020scalinglawsneurallanguage}
Jared Kaplan, Sam McCandlish, Tom Henighan, Tom~B. Brown, Benjamin Chess, Rewon
  Child, Scott Gray, Alec Radford, Jeffrey Wu, and Dario Amodei.
\newblock Scaling laws for neural language models, 2020.

\bibitem{kingma2022autoencodingvariationalbayes}
Diederik~P Kingma and Max Welling.
\newblock Auto-encoding variational bayes, 2022.

\bibitem{kolchinsky2024thermodynamicdissipationdoesbound}
Artemy Kolchinsky.
\newblock Thermodynamic dissipation does not bound replicator growth and decay
  rates, 2024.

\bibitem{10.1098/rstb.2009.0295}
Leonid~M. Martyushev.
\newblock The maximum entropy production principle: two basic questions.
\newblock {\em Philosophical Transactions of the Royal Society B: Biological
  Sciences}, 365(1545):1333--1334, 05 2010.

\bibitem{10.3389/fmicb.2017.02019}
Volker M^^c3^^bcller and Verena Hess.
\newblock The minimum biological energy quantum.
\newblock {\em Frontiers in Microbiology}, Volume 8 - 2017, 2017.

\bibitem{nagayama2026dualitydissipationcoherencetradeoffthermodynamic}
Ryuna Nagayama and Sosuke Ito.
\newblock Duality between dissipation-coherence trade-off and thermodynamic
  speed limit based on thermodynamic uncertainty relation for stochastic limit
  cycles, 2026.

\bibitem{PhysRevE.101.022415}
William~D. Pi\~neros and Tsvi Tlusty.
\newblock Kinetic proofreading and the limits of thermodynamic uncertainty.
\newblock {\em Phys. Rev. E}, 101:022415, Feb 2020.

\bibitem{PhysRevLett.119.240601}
Matteo Polettini and Massimiliano Esposito.
\newblock Effective thermodynamics for a marginal observer.
\newblock {\em Phys. Rev. Lett.}, 119:240601, Dec 2017.

\bibitem{e27040449}
Yasuji Sawada, Yasukazu Daigaku, and Kenji Toma.
\newblock Maximum entropy production principle of thermodynamics for the birth
  and evolution of life.
\newblock {\em Entropy}, 27(4), 2025.

\bibitem{WhatisLife}
E.~Schr^^c3^^b6dinger.
\newblock {\em What is Life?}
\newblock Cambridge University Press., 1944.

\bibitem{shazeer2017outrageouslylargeneuralnetworks}
Noam Shazeer, Azalia Mirhoseini, Krzysztof Maziarz, Andy Davis, Quoc Le,
  Geoffrey Hinton, and Jeff Dean.
\newblock Outrageously large neural networks: The sparsely-gated
  mixture-of-experts layer, 2017.

\bibitem{PhysRevLett.121.070601}
Naoto Shiraishi, Ken Funo, and Keiji Saito.
\newblock Speed limit for classical stochastic processes.
\newblock {\em Phys. Rev. Lett.}, 121:070601, Aug 2018.

\bibitem{smith2017cyclicallearningratestraining}
Leslie~N. Smith.
\newblock Cyclical learning rates for training neural networks, 2017.

\bibitem{NEURIPS2019_3001ef25}
Yang Song and Stefano Ermon.
\newblock Generative modeling by estimating gradients of the data distribution.
\newblock In H.~Wallach, H.~Larochelle, A.~Beygelzimer, F.~d\textquotesingle
  Alch\'{e}-Buc, E.~Fox, and R.~Garnett, editors, {\em Advances in Neural
  Information Processing Systems}, volume~32. Curran Associates, Inc., 2019.

\end{thebibliography}
\end{document}